\documentclass[multphys,vecphys]{svmult}

\usepackage{makeidx}         % allows index generation
\usepackage{graphicx}        % standard LaTeX graphics tool
\usepackage{multicol}        % used for the two-column index
\usepackage[bottom]{footmisc}% places footnotes at page bottom

\makeindex             % used for the subject index
\begin{document}

\title*{Past and Future of CG J1720-67.8: Constraints from
Observations and Models}
\titlerunning{Past and Future of CG J1720-67.8}
\author{Sonia Temporin\inst{1,2}\and
Wolfgang Kapferer\inst{1}}
\institute{Institut f\"ur Astro- und Teilchenphysik, 
Leopold-Franzens University Innsbruck, Technikerstra\ss e 25, A-6020
Innsbruck\\
\texttt{giovanna.temporin@uibk.ac.at, wolfgang.e.kapferer@uibk.ac.at}
\and INAF - Brera Astronomical Observatory, Via Brera 28, I-20121, Milano}
\maketitle

\begin{abstract}
We discuss the evolution of the peculiar, nearby (z = 0.045), compact
galaxy group CG J1720-67.8, by interpreting a large amount of
observational information on the basis of our recent results 
from spectrophotometric evolutionary synthesis models and new 
N-body/SPH simulations. The group, that is composed of two spiral galaxies with
a mass ratio approximately 4:1 and an S0 galaxy in a particularly compact
configuration, is undergoing an active pre-merging phase.
Several tidal features
%, among which a prominent star-forming tidal tail hosting some
%candidate tidal dwarf galaxies, 
are signposts of the complex dynamics
of the system. We suggest that the observed structure of the tidal features
can be explained only if all three galaxies are involved in a strong interaction
process.
\end{abstract}

\section{Introduction}
\label{temp:sec1}
Several catalogues of compact galaxy groups (CGs) have been compiled
during the last 20 years (e.g. V. Eke, this volume). Many studies
have been/are devoted to understand the nature of CGs and their evolution
from both an observational and a theoretical point of view (see e.g. 
the contributions by J. Hibbard, G. Mamon, E. Pompei, this volume).
Nevertheless, we do not have yet a clear understanding of how groups evolve
and what the product of their evolution is. The study of individual groups
in differing evolutionary phases might be useful in this respect, in
particular for rarely observed phases, like the pre-merging state represented
by extremely compact systems.
This is the main motivation of our detailed studies of such a compact
system, CG J1720-67.8. In fact, this group gives us the rare opportunity 
to gather information on the evolutionary stage
that precedes the coalescence and provides us with
an indication of the processes that might be at play at higher redshifts.
In the last few years we collected a large amount of
observational information on this system to disentangle its evolutionary
history. The application of evolutionary synthesis models allowed us to
date the latest bursts of star formation, that are related to the most
recent galaxy encounters. In the following we summarize some of the most significant
results concerning this system, and discuss its evolution on the basis of 
the comparison -- presented here for the first time -- with N-body/hydrodynamical 
simulations.

\section{A very evolved galaxy group}
\label{temp:sec2}

\begin{figure}
\centering
%\hbox{
%\includegraphics[height=3.6cm]{temporinF1a.eps}
%\includegraphics[height=3.6cm]{temporinF1b.eps}}
\includegraphics[height=3.6cm]{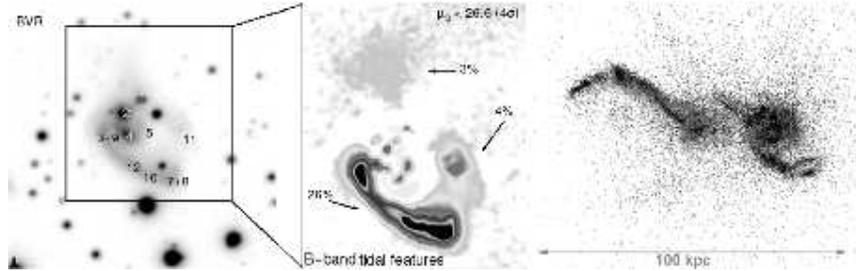}
\caption{\emph{Left.} Composite BVR image of CG J1720-67.8. The galaxies and tidal dwarf candidates
are labeled.  \emph{Middle.} Zoom on the tidal features 
on the B image after subtraction of the stars and bi-dimensional galaxy models. \emph{Right.} A timestep 
of the simulation of a prograde encounter between two unequal mass galaxies. The stellar component
is plotted in grey and the gaseous component in black (see Sect.~\ref{temp:sec4}).}
\label{temp:fig1}       % Give a unique label
\end{figure}
The group CG J1720-67.8 \cite{temp:t03a,temp:t03b}, composed of two spiral galaxies (G1 and G4),
one S0 galaxy (G2), and a number of tidal dwarf galaxy 
candidates in a very compact configuration (Fig.~\ref{temp:fig1}), appears to be next to
the final merging \cite{temp:t05}. All components show significant star formation 
activity. In fact the group has a total (uncorrected) H$\alpha$ luminosity of 
3.7 $\times$ 10$^{41}$ ergs s$^{-1}$ and the total star formation rate derived from the 
20 cm radio continuum emission -- which is largely unaffected by extinction -- is 
18 M$_{\odot}$ yr$^{-1}$.  
The prominent tidal features, which account for about one third of the optical luminosity and 
of the H$\alpha$ emission of the group, indicate that the galaxies have recently undergone 
violent interactions. The age of the most prominent tidal tail, based on its projected length and
on the kinematics of G4, is estimated to be approximately 200 Myr. This is indicative
of the age of the latest strong interaction event.  

The velocity field of the group (Fig.~\ref{temp:fig3}) obtained from the H$\alpha$ emission
line gives evidence of its complex kinematics. The kinematic center is offset from the center of 
G4 by about 5 kpc.
Galaxies G1 and G4, and the bridge of ionized gas between them have similar systemic radial velocities.
Radial velocities gradually increase along the tidal tail from south to north, with a strong
gradient in the northern part, near G4 and G2. There is no visible sign of return
of tidal material to the parent galaxy. 
The deficiency of neutral hydrogen in compact groups has been suggested as an indicator of
their evolutionary state \cite{temp:verd01}. The lack of detected neutral hydrogen in
CG J1720-67.8, which gives an upper limit to the HI content of
2.3 $\times$ 10$^9$ M$_{\odot}$, would suggest that the group is HI-deficient, if compared
with the total B luminosity \cite{temp:t05}. However, in this case the B-band luminosity
is particularly enhanced due to the star formation activity, hence it leads to an overestimate of
the expected neutral hydrogen content of the group. The small size of the galaxies 
(diameters of $\approx$ 8 and 14 kpc for G1 and G4), following \cite{temp:hg84}, suggest an expected 
HI mass $\leq$ 2 $\times$ 10$^9$ M$_{\odot}$, comparable to our estimated upper limit.
The diameters of these strongly interacting galaxies are ill-defined, however, 
the whole galaxy group with its tidal features, has an optical size comparable to that of our
Galaxy, so it is expected to have a similar content of neutral gas. As a consequence, in this case 
the available upper limit to the HI mass is not sufficient to establish whether the group is 
HI-deficient or not.

\subsection{Results from evolutionary synthesis models}
\label{temp:sec3}

The application of chemically-consistent spectrophotometric evolutionary synthesis models to the
three main galaxies of CG J1720-67.8 provided us with information about the
age and strength of the latest interaction-induced bursts of star formation \cite{temp:t06}. 
According to the best-fit models, the two spiral galaxies underwent a strong burst of star formation
40 -- 180 Myr ago. An older burst of star formation in the S0 galaxy accompanied 
either a merger event or an interaction with the companions, $\leq$ 1 Gyr ago.
The total (stars + gas) masses of the galaxies implied by the best-fit models are 
3.4 -- 7 $\times$ 10$^{10}$ M$_{\odot}$ (depending on the considered model) for G2,
4 -- 6 $\times$ 10$^{9}$ M$_{\odot}$ for G1, and 2 $\times$ 10$^{10}$ M$_{\odot}$ for G4.
Therefore, the total stellar plus gaseous mass of the galaxies involved, without taking
into account the material in the tidal tail(s), is $\geq$ 6 $\times$ 10$^{10}$ M$_{\odot}$.
The visible mass of the final group remnant is expected to be on the order 10$^{11}$ 
M$_{\odot}$.

\section{Clues from hydrodynamical simulations}
\label{temp:sec4}

\begin{figure}
\centering
\includegraphics[height=4.2cm]{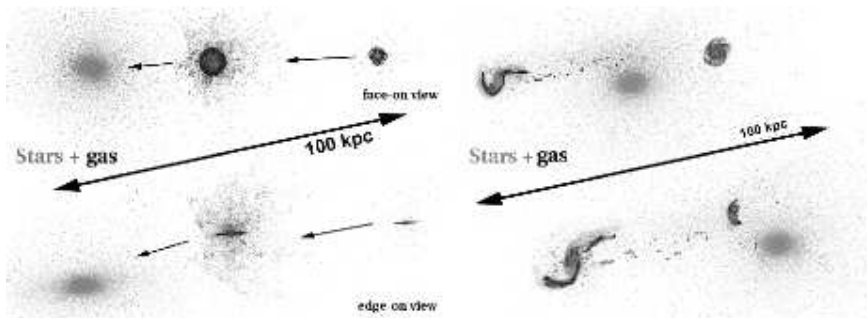}
\caption{\emph{Left.} N-body/SPH simulation of the encounter between two disks and an early-type galaxy.
At this time-step the larger spiral has started the interaction. \emph{Right.} Subsequent timesteps 
of the simulation after the encounter of the S0 with the larger spiral (top) and with the smaller spiral
(bottom).}
\label{temp:fig2}       % Give a unique label
\end{figure}

\begin{figure}
\centering
\includegraphics[height=5.0cm]{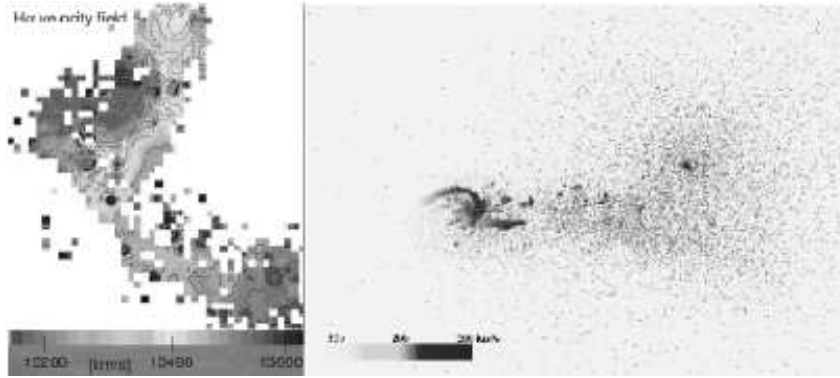}
\caption{\emph{Left.} Isovelocity contours overlaid to the velocity field of the gaseous component (from
\cite{temp:t05}). \emph{Right.} Projected radial velocity distribution of the gas in the N-body/SPH 
simulation at the last timestep shown in Fig.~\ref{temp:fig2}.}
\label{temp:fig3}       % Give a unique label
\end{figure}

Observations alone are not sufficient to establish the interaction
history of the group, although they give important indications. 
Extensive numerical simulations are required
for this purpose. Here we present a first exploration of the possible 
evolutionary paths that could explain the observed galaxy 
configuration, on the basis of combined N-body/SPH simulations.
For the modelling we used GADGET-2 \cite{temp:vs05}. For general 
assumptions and technical details of the simulations we refer the reader
to \cite{temp:kapf05} and references therein.
Since the two spiral galaxies G1 and G4 have similar systemic velocities
and appear to be connected by a bridge of ionized gas, we first
compared the observations with an encounter between two unequal mass disks.
The best-fit evolutionary synthesis models imply for these two galaxies a mass ratio
$\approx$ 1:3 or 1:4, hence a 1:3 mass ratio was adopted in the simulations.
The properties of the simulated galaxies are similar to those listed in
table~1 of \cite{temp:kapf05}.
Both fast prograde encounters and retrograde encounters between the two
disks produce a knotty tidal tail departing from the more
massive galaxy, but also a relatively prominent counter-tail departing
from the less massive galaxy (see e.g., the timestep shown in Fig~\ref{temp:fig1}). 
The latter is not observed in CG J1720-67.8.
This points to a non-negligible role of the S0 galaxy in the interaction.
The presence of the third galaxy could suppress the formation of the counter-tail
in the smaller spiral. To test this hypothesis, we introduced the S0 galaxy in our 
simulation. The formation of the S0 galaxy was simulated through an equal mass disk 
merger. After 1 Gyr a massive disk approaches the S0 and starts a burst of star
formation accompanied by galactic winds (Fig.~\ref{temp:fig2}, left). 
The encounter of this spiral with the merger remnant
produces a knotty tidal arm and a bridge of stars and gas toward the S0 (Fig.~\ref{temp:fig2}, 
top-right). Another 15 Myr
later, in turn, the less massive disk undergoes an encounter with the S0,
is strongly distorted by it, without producing any significant tidal tail (Fig.~\ref{temp:fig2}, 
bottom-right), and starts an induced burst of star formation. 
Some residual star formation activity is still present at the center of the S0.
Although this simulation thus not reproduce the galaxy configuration of CG J1720-67.8, it shows several 
features similar to the observed ones concerning the morphological distortions, the tidal tails,
and the induced star formation. Also the velocity field, projected along a hypothetical line-of-sight,
has some similarities with the observed one, as shown in Fig.~\ref{temp:fig3}.

\vspace{-0.5in}

\section{Conclusions}
\label{temp:sec5}

We found several observational evidences that CG J1720-67.8 is in a late pre-merging phase.
The evolutionary stage of the group is similar to that of 
HCG 31, which is claimed to have started its merging process \cite{temp:amr04}.
According to evolutionary synthesis models the latest interaction 
episode for the two spiral galaxies took place $<$ 200 Myr ago. Another interaction
or merger event appears to have involved the S0 galaxy less than 1 Gyr ago.
Our first comparisons with N-body/hydrodynamical simulations indicate that the observed 
tidal features cannot be simply the result of the interaction between
the two spiral galaxies. Therefore, we suggest that all 
three galaxies have been involved in the latest interaction. 
The extreme compactness and low radial velocity dispersion suggest that the group 
coalescence will be fast.
A simulation in which an S0 galaxy is formed through an equal mass merger and about 1 Gyr 
later undergoes an encounter with both spiral galaxies produces results that approach the 
observed properties of the group. 
The properties of CG J1720-67.8 suggest that sufficiently gas-rich groups might undergo
a particularly active star-forming phase before final coalescence. 

\subparagraph{Acknowledgments.} ST is grateful to R. Giovanelli for useful discussions 
during the meeting. The authors acknowledge financial support by the Austrian Science 
Fund (FWF) under projects P17772 and P15868.%

%\input{referenc}
%%%%%%%%%%%%%%%%%%%%%%%%%%%%%%%%%%%%%%%%%%%%%%%%%%%%%%%%%%%%%%%%%%%%%%  }

%%%%%%%%%%%%%%%%%%%%%%%%%%%%%%%%%%%%%%%%%%%%%%%%%%%%%%%%%%%%%%%%%%%%%%

\printindex
\end{document}